# Construction of a Single-Column Model in RegCM4 and its Preliminary Application for Evaluating PBL Schemes in Simulating the Dry Convection Boundary Layer

Zhenyu HAN

National Climate Center, China Meteorological Administration, Beijing 100081, China

**Corresponding author**: Zhenyu HAN

**Corresponding address**: National Climate Center, China Meteorological Administration, Beijing 100081, China

**E-mail**: hanzy@cma.gov.cn

**ABSTRACT**

A single-column model (SCM) is constructed in the regional climate model RegCM4. The evolution of a dry convection boundary layer (DCBL) is used to evaluate this SCM and compare four planetary boundary layer (PBL) schemes, the Holtslag-Boville scheme (HB), Yonsei University scheme (YSU), and two University of Washington schemes (UW01, Grenier-Bretherton-McCaa scheme and UW09, Bretherton-Park scheme), using the SCM approach. A large-eddy simulation (LES) of the DCBL is performed as a benchmark to examine how well a PBL parameterization scheme reproduces the LES results, and several diagnostic outputs are compared to evaluate the schemes. In general, with the DCBL case, the YSU scheme performs best for reproducing the LES results, which include well-mixed features and vertical sensible heat fluxes; UW09 has the second best performance, UW01 has the third best performance, and the HB scheme has the worst performance. The results show that the SCM is proper constructed. Although more cases and further testing are required, these simulations show encouraging results towards the use of this SCM framework for studying the physical processes in RegCM4.

## 1. Introduction

Although climate numerical models have had a great development in recent years, many physical processes such as turbulent and diffusion processes in the planetary boundary layer (PBL) still cannot be fully resolved partly due to coarse resolution. Therefore, physical parameterization is indispensable and critical in these models, and parameterization testing is a vital task in the process of model development. The easiest and most widely used approach is by application of climate simulations, the results of which can be directly compared with multiple observations or reanalysis datasets. However, one disadvantage is that it can be very difficult to attribute simulation deficiencies to particular aspects of a model's formulation because various feedbacks, such as the interplay between dynamics and physics, are mingled together during model integration (Wang 2015). The single-column model (SCM) is an economical framework for developing and diagnosing the physical processes in climate models, and with this tool, a parameterization can be tested by evaluating its ability to reproduce the observed tendencies for a given large scale situation.

Several regional climate modeling (RCM) or limited area modeling (LAM) groups have constructed SCMs (Table 1); however, there has been no reported SCM construction for the regional climate model RegCM4 until now (Giorgi et al. 2012) (The ARCSyM is an Arctic version of RegCM2, which has an SCM called ARCSCM and has not been integrated into the RegCM4's released versions; Morrison et al. 2003). In this study, an SCM is developed with most of the parameterizations inherited from the RegCM4. For ease of construction and use, this SCM is designed exactly following the framework of the 3D RegCM4.

There is no absolute best PBL parameterization scheme, because each schemes have both advantages and disadvantages contributing to various assumptions and formulations (e.g., Cohen et al. 2015; Wang et al. 2016). So a deep understanding of the physical behavior of PBL schemes will help improve PBL parameterizations and interpret simulation deficiencies. Then, the constructed model in this study is used in sensitivity studies of SCM simulations for the PBL schemes.

To assess the performance of a PBL scheme, well-controlled cases are usually used to isolate the contribution of PBL processes, which are either ideal cases or simplified real cases. A previous evaluation of PBL schemes in the SCM framework of the Community Atmosphere Model (CAM) by Bretherton and Park (2009) focused on three types of PBLs: (1) the dry convection boundary layer (DCBL); (2) stably stratified boundary layer; and (3) nocturnal stratocumulus-topped boundary layer, which have also been widely used as testbeds in past intercomparison studies. The first purpose of this study is to test if the SCM has been correctly constructed, and thus, the most fundamental case of the DCBL is chosen. Based on this case, the basic performances of the four different PBL schemes are also evalulated.

The large-eddy simulation (LES) models can robustly reproduce observed DCBLs without significant model dependence, and have been widely used as benchmarks (e.g., Noh et al. 2003; Bretherton and Park 2009; Shin and Dudhia 2016; Wang et al. 2016). We first simulated the evolution of a DCBL using a LES model.

Then, the SCM with each PBL parameterization scheme is driven by the same prescribed surface heat fluxes and initial conditions as those of the LES run. The PBL characteristics simulated from the SCM runs are compared with each other and with those derived from the LES data.

In Section 2, the DCBL simulation sets of both the LES and SCM, construction of the SCM, and brief summary of the PBL schemes used in this study are described. The evaluations of the simulated PBL features from the SCM are presented in Section 3, and in Section 4, a summary is provided.

## 2. Model, Experimental Design, and Method

### 2.1 LES Benchmark Simulation

The University of California, Los Angeles, large-eddy simulation (UCLA-LES; Stevens et al. 2005) model is used to simulate a DCBL explicitly. The PBL flow is driven by the prescribed surface sensible heat flux of 300 W $m^{-2}$, and the surface temperature is derived based on the flux-gradient relation, with the roughness is set to 0.1 m . The initial profiles are set with a potential temperature of $\theta$ = 288 K + (3 K $km^{-1}$) $\times z$ and the wind component values are $u$ = 10 m $s^{-1}$ and $v$ = 0 m $s^{-1}$. The surface pressure is set to 1000 hPa within the whole simulation period. The Coriolis acceleration is turned off, and there is no moisture, large-scale vertical motion, or radiative heating.

The resolution set uses the typical configuration, including that the horizontal extent covers the domain of 10 × 10 $km^2$ with 50-m resolution; the vertical extent reaches a height of 5 km with 20-m resolution; a sponge layer occupies the upper ten levels. A 16-h-long simulation is conducted, with the first hour being excluded from the analysis as model spin up. The instantaneous fields over time and over different heights are averaged to derive 5-min and hourly mean variables.

**2.2 SCM Model Construction**

Most parts of the SCM, including the dynamic core and physics packages are the same as those of the 3D RegCM4 (model version v4.4). To minimize changes in the original codes (e.g., staggered Arakawa B-grid), a 4 × 4 grid but not a single vertical column is set as the dynamic core of the SCM. On this 4 × 4 grid, all horizontal dynamical processes (horizontal diffusion and advection) and lateral boundary conditions are turned off, but the cyclic boundary conditions in both the x and y directions are added, and thus, the values of all variables are the same among those grid points. Then, any point from this 4 × 4 grid can be considered a "single column".

As with other SCMs (e.g., SCMs in WRF and CAM), horizontal temperature and moisture advective tendencies, as well as vertical velocity or vertical advection, can be prescribed as inputs to drive the SCM. However, all these modules are switched off in this study.

**2.3 SCM Simulations**

To ensure that the discrepancies in the simulated PBL flow are only due to differences in the PBL schemes, only the PBL and surface layer parameterization along with dry dynamical core are activated in the DCBL simulation run. The four PBL schemes used in this paper are the Holtslag-Boville scheme (HB; Holtslag et al. 1990; Holtslag and Boville 1993), Yonsei University scheme (YSU; Hong et al. 2006;

Hong 2010), and two University of Washington schemes (UW01 and UW09). The UW01 is based on Grenier and Bretherton (2001) and Bretherton et al. (2004), while the UW09 is based on Bretherton and Park (2009). The HB has been part of the RegCM series models since the early version, the UW01 was added to the RegCM4 by O'Brien et al. (2012), and the YSU and UW09 were added by the author of this study, and the codes are modified from the WRF v3.5.1 and CESM v1.2.0 models, respectively.

The HB and YSU are the non-local, first-order closure schemes in which the diffusion coefficient profile is an empirical function of both the surface fluxes and fractional height within the boundary layer. The turbulence variables are diagnosed based on the diffusion coefficient, local gradient, and a non-local gradient correction term. The major difference between the two schemes is that the entrainment processes is explicitly considered in the YSU.

The UW01 and UW09 are the local, 1.5-order closure schemes in which the turbulent kinetic energy (TKE) is predicted or diagnosed, and other turbulence variables are diagnosed based on the local TKE. The major difference between the two schemes is the calculation method of the TKE. For more details on the four PBL schemes, please refer to the references.

In the 3D RegCM4, the surface layer scheme is imbedded in the land surface model. Because the land surface model is not activated in this study, a simple surface layer scheme is added in the SCM, which is extracted from the BATS land model (Dickinson et al. 1993). With this surface layer scheme, given the prescribed heat fluxes of 300 W $m^{-2}$ and the calculated surface temperature from the LES simulation, the surface bulk Richardson number, drag coefficents, and fractional velocity can all be derived. Other surface parameters are set to the same value as in the LES model runs.

A stretched vertical coordinate is used such that finer spacing is assigned to the lower levels while coarser vertical spacing is applied at higher levels. The vertical resolution set in the control SCM run is the default 18-level set of 3D RegCM4, with the model top set at 50hPa. The vertical grid size is approximately 80 m near the surface and 900 m near the 3 km height above the surface. In this set, the SCM runs are called HB, YSU, UW01, and UW09. Another vertical resolution set is the 41 level, which is used to detect how the vertical resolution affects on the simulations, with a vertical grid size of approximately 80 m near the surface and 250 m near the 3 km height above the surface, and with the model top set at 50hPa. In this set, the SCM runs are called HB_z41, YSU_z41, UW01_z41, and UW09_z41, which is in contrast with the 18 level runs. All initial conditions and the model integration set in the SCM runs are the same as those in the LES.

### 2.4 Diagnostic Output

1) PBL height ($Z_{PBL}$), depth of mixed layer ($H_{ML}$), and mixing index (MI)

Determining the PBL height ($Z_{PBL}$) is important in atmospheric numerical models because $Z_{PBL}$ is used in both the PBL scheme itself (e.g., to scale the eddy diffusivity in the HB and YSU scheme) and in other physical parameterizations where required (e.g., to scale the strength of the convective velocity scale used in the wind speed component of the sea surface fluxes (Zeng et al. 1998)).

All four PBL schemes and the LES model provide the PBL heights as part of their output variables, but the computation methods are not coherent among the schemes and the LES. Since the calculation method in a particular PBL scheme is a characteristic of the scheme, we first analyze the diagnosed PBL height directly from the five experiments ($Z^0_{PBL}$). Because $Z^0_{PBL}$ depends on the diagnosis method used in different PBL schemes and LES (e.g., LeMone et al. 2013; Wang et al. 2016), during post processing the unified diagnosis method is added to derive the re-diagnosed PBL height ($Z^1_{PBL}$) of all PBL schemes and LES for comparison. The bulk Richardson number method is applied to re-diagnose the PBL height using data from all SCM and LES model simulations. In this method, the PBL height is set as the height *z* when bulk Richardson number between *z* and surface is equal to 0.25. With this method, the $Z^1_{PBL}$ is not restricted to the model levels, indicating that it is not very sensitive to the distribution and resolution of the vertical layers, especially in lower vertical resolution cases.

Following Wang et al. (2016), two extra variables are calculated, which describe the uniformity of a mixed PBL, the thickness of the well-mixed layer ($H_{ML}$) and mixing index (MI). A well-mixed layer is defined as the layer with a very small vertical gradient (the absolute value less than 0.20 K km$^{-1}$) of potential temperature. During the calculation of $H_{ML}$, the top and bottom of the well-mixed layer is not restricted to the vertical levels but can be intepolated between levels. The MI is measured by the standard vertical deviation of potential temperature within the well-mixed layer divided by the $H_{ML}$, then multipled by 10 to make the value more readable.

2) Vertical fluxes and entrainment flux

The vertical fluxes of sensible heat could not be obtained from the PBL schemes in the SCM directly. For comparison purposes, the vertical sensible heat flux $\langle w\,\theta \rangle_z$ at a certain height ($z$) is calculated by integrating the PBL $\theta$ tendency $\left(\frac{\partial \theta}{\partial t}\right)_{PBL}$ from the surface to height $z$, which is as follows:

$$\langle w\,\theta \rangle_z = \langle w\,\theta \rangle_{z=0} - \int_0^z \left(\frac{\partial \theta}{\partial t}\right)_{PBL} dz \qquad (1)$$

where $\langle w\,\theta \rangle_{z=0}$ is the surface sensible heat flux. The entrainment flux of the sensible heat is estimated as the minimum sensible heat flux near the PBL top, and the entrainment zone is the layer with a negative sensible heat flux. As mentioned in Wang et al. (2016), a disadvantage of this derivation method is the accumulation of numerical errors during the vertical integration, but these errors within the PBL are quite small in our study.

**3. Results**

The LES simulation shows that the frictional velocity, $u_*$, decreases over time (Fig. 1a). All four SCM experiments in the 18-level sets can simulate the changes in $u_*$; however, the SCM simulated $u_*$ is slightly smaller than the LES simulated result with biases of -0.04 m/s ~ 0.01 m/s. The simulated $u_*$ values from the UW09 are closest to the LES results (Fig. 1a). Generally, the difference between the two vertical resolution sets is very small. Compared to the 18-level runs, the curves of $u_*$ in the 41-level runs are smoother and the discrepancies among the HB, YSU, and UW01 are much smaller; however, the magnitudes show little change (Fig. 1b). Overall, the well-simulated $u_*$ indicates that the module of the surface layer processes has been correctly constructed.

**3.1 PBL Height**

Figure 2a shows the diagnostic output, $Z_{PBL}^{0}$, which is directly from the respective PBL schemes with 18-level set and the LES run. The top of PBL is raised continuously due to the persistent surface heating during the simulation. In general, the time evolution of PBL height is well reproduced by the SCM simulations using all schemes. However, the magnitudes and smoothness of the curves are quite different among the four schemes. In the HB and YSU schemes, the PBL height for unstable conditions is determined to be the first neutral level by checking the bulk Richardson number, which is calculated between the lowest model level and the levels above. This approach permits the PBL top to lie between model levels and evolve continuously over time. In the LES run, the PBL height is defined by the height of the maximum potential temperature gradient, which has a time series that is also quite smooth due to the very high vertical resolution. However, in the UW01 and UW09 schemes, the height is restricted to lie on the model levels, and thus, the time evolution is not continuous (can also be seen in Fig. 4 of Grenier and Bretherton (2001)).

After re-diagnosed using the same methods, the $Z_{PBL}^{1}$ in the SCM is more consistent with that in the LES, for both the time evolution and magnitude (Fig. 2c). All SCM results overestimate the PBL height, and the bias from the HB scheme is the largest. The higher vertical resolution does not change much, the curves of $Z^0_{PBL}$ and $Z^1_{PBL}$ are smoother, and the HB scheme is still the worst in the 41-level runs (Figs. 2b and 2d).

### 3.2 Wind and Temperature

Figures 3a and 3b show comparisons of the vertical profiles of the hourly mean wind speeds at 5 h and 9 h from all 18-level SCM simulations. As shown by the LES simulation at 5 h in Fig. 3a, the wind speed near the surface and PBL top increases with height due to surface drag and entrainment, respectively, while the speed within the mixed layer is nearly constant due to being well mixed. At 9 h, with the PBL top rising, the mean wind speed within the well-mixed PBL decreases over time due to the synergic effect of surface drag and PBL mixing (Fig. 3b). These features are well simulated by the UW01, UW09, and YSU experiments, and the profiles are similar among all three runs. However, the simulated wind speed within the mixed layer from the HB experiment is not well mixed, and there is a large vertical gradient, because the non-local gradient correction term is not included in the momentum prognostic equation (Güttler et al. 2014).

Figure 3c shows the wind profiles after raising the vertical resolution. There are larger differences among the four SCM-simulated wind profiles when compared to the lower resolution runs. Raising the vertical resolution in the lower atmosphere improves the simulation of the wind profiles for the YSU and UW09 schemes. The YSU_z41 produces the best simulation among the four schemes, while in the UW09_z41, the wind speed simulated below the boundary layer top is still slightly underestimated. For the UW01 and HB schemes, the use of a higher vertical resolution does not change much. The entrainment zone is much lower in the UW01_z41 and much higher in the HB_z41 compared to that in the LES result, and the HB is still the worst scheme, as the vertical gradient bias is not reduced much.

Figures 4a and 4b show the hourly mean potential temperature profiles at 5 h and 9 h from all SCM simulations with the 18-level set. As shown by the LES simulation at 5 h in Fig. 4a, there is a unstable layer in the lower part of the PBL, a well-mixed layer with small potential temperature gradient in the mid-PBL, and a stable layer in the upper part of the PBL. At 9 h, with the PBL top rising, the mean potential temperature within the mid-PBL increases (Fig. 4b). The main discrepancies from different schemes with the 18-level set lie in the thickness of the well-mixed layer and potential temperature gradient within the well-mixed layer (Figs. 4a and 4b). Generally, the UW01, UW09, and YSU simulations are similar to each other, and the HB simulations produce the largest discrepancy relative to the LES results. In the HB, the potential temperature gradient within the mixed layer is largely overestimated, and there is a weak inversion layer between the surface layer and mixed layer, which is more intense at 5 h.

Raising the vertical resolution in the lower atmosphere aids simulation of the vertical structures in the potential temperature profiles (Fig. 4c). Both the YSU and UW09 can produce nearly the same vertical profiles as that of the LES. For the UW01 and HB schemes, there are still large biases. Compared with the LES results, the PBL is more unstable and shallower in the UW01_z41 and more stable and deeper in the HB_z41. In addition, the temperature of the lower atmosphere in the UW01 scheme tends to be colder, and the temperature in the HB scheme tends to be warmer, which may correspond to the warm bias reduction in the long-term climate simulation when the PBL scheme is changed from the HB to the UW01 (Gütler et al. 2014). There is still a fake inversion layer between the surface layer and mixed layer in the HB_z41, which is more intense than that in the lower vertical resolution simulation and exists all time (figures do not show). This fake inversion layer is due to a deficient paramerization of the eddy diffusivity in the HB scheme, which will be further discussed in the following paragraphs.

Table 2 shows a comparison of the $H_{ML}$ and MI values from all simulations at 5 h and 9 h, which are calculated based on the hourly mean profiles. There are large biases in these two variables from the SCM results with the 18-level set, which is partly due to the low vertical resolution, because most of the biases are reduced after raising the vertical resolution. In both SCM runs with 18- and 41-level sets, the $H_{ML}$ and MI values from the YSU are closest to those from the LES, indicating a more uniformaly mixed PBL among those SCM results. This could be attributed to both the non-local mixing and entrainment parameterization in the YSU scheme. The UW09 is also well mixed, and the bias is greatly reduced after raising the vertical resolution. The biases in the HB and UW01 schemes with the 41-level run are still large, which is consistent with the conclusion from the profile evalutions.

### 3.3 Flux of Sensible Heat

Figure 5a and 5b show hourly mean vertical fluxes of sensible heat at 5 h and 9 h with the 18-level SCM and LES, expressed as the ratio of vertical flux and surface flux. In the LES model, the ratio of PBL top entrainment flux and surface flux is approximately −0.2, which is consistent with lots of previous studies, indicating the entrainment flux is about -60 W $m^{-2}$ (-0.2 × 300 W $m^{-2}$). It shows that the UW01, UW09, and YSU produce less downward entrainment buoyancy flux at the PBL top, while the HB scheme produces more flux. Among the schemes, the value from the YSU scheme is closest to the LES value. All schemes overestimate the height of the minimum buoyancy flux. Due in part to the numerical error from the integral calculation of the vertical flux in the SCM, the flux value cannot be quickly reduced to around zero above the top of PBL. These features can also be clearly seen from the time evolution figure (Figs. 6a, 6c, 6e, and 6g). The $Z^0_{PBL}$ is close to the height of the minimum buoyancy flux, especially in the UW01 and UW09 schemes, and it helps to indicate the time evolution of the entrainment zone height. Overestimates of the entrainment flux at the PBL top always exist in the HB scheme (Fig. 6a).

After raising the vertical resolution, the entrainment zones are better resolved in all schemes (Fig. 5c). The YSU remains the scheme with the lowest bias of heat flux, while the HB_z41 overestimates the entrainment flux and height of the entrainment zone, and the UW01_z41 and UW09_z41 underestimate both. As shown in the time evolution figure, the biases of the heat flux change little over time (Figs. 6b, 6d, 6f, and 6h).

**3.4 Eddy Diffusivity and TKE**

Figure 7a shows the eddy diffusivity profiles for heat ($K_h$) after 5 h with the 18-level set. All results have a 5-min average. This result shows that the difference in the $K_h$ magnitude among the schemes is very large. The largest diffusivities appear in the UW09 scheme with a vertical maximum $K_h \approx 1500$ m$^2$, and the maximum $K_h$ values vary from 500 to 900 m$^2$ s$^{-1}$ in other schemes. The $K_h$ profiles shown here only characterize the local mixing ability in the HB and YSU schemes, because other part of the turbulent mixing in these schemes is also represented by their non-local mixing treatments. So the shapes of $K_h$ in the HB and YSU are quite different from those in the UW01 and UW09 schemes, and the location of the maximum diffusivity values are lower. The differences in the shape are more obvious in the 41-level set (Fig. 7b).

The time evolution figures (Figs. 8a, 8c, 8e, and 8g) show that as the PBL top raises, the maximum values of $K_h$ generally increase over time, and diffusivities larger than 10 m$^2$ s$^{-1}$ also extend to higher levels. Diffusivity profiles are limited below $Z_{PBL}$ in all schemes, because the $K_h$ profiles are parameterized as so, although the detailed formulations of $K_h$ are different among the four schemes. Compared with the smooth evolutionary features at high resolution (Figs. 8b, 8d, 8f, and 8h), the low resolution results show a fluctuating evolution, synchronizing with the change in $Z_{PBL}$.

Figures 7c and 7d present the vertical distribution of the Prandtl number, where $Pr = K_m/K_h$. In the surface layer and mixed layer, the Pr values from all schemes except the YSU are nearly constant and smaller than 1.0, while the Pr value in the YSU increases upward. Above the mixed layer, in both the HB and UW09 schemes, the Pr decreases to 1.0; and in the UW01 scheme, the Pr decreases to a constant value larger than 1.0; while in the YSU scheme, the Pr profile is quite different, showing that within the entrainment zone the Pr increases beyond 1.0, but above the PBL top the Pr decreases to 1.0 quickly.

Near the surface layer top, there is significant discontinuity on the Pr in the HB scheme due to different Pr equations being used between the surface layer and mixed layer as follows:

$$\mathrm{Pr} = \mathrm{Pr}_{ML} = \frac{\phi_h}{\phi_m}\left(\frac{0.1 * Z_{PBL}}{L}\right) + a * k * \frac{0.1 * Z_{PBL}}{Z_{PBL}}$$

$$= \left(1 - 15 * \frac{0.1 * Z_{PBL}}{L}\right)^{-\frac{1}{6}} + 0.34,$$

$$\text{, when } z \geq 0.1 * Z_{PBL}$$

$$\mathrm{Pr} = \frac{\phi_h}{\phi_m}\left(\frac{z}{L}\right) = \left(1 - 15 * \frac{z}{L}\right)^{-\frac{1}{6}},$$

$$\text{, when } z < 0.1 * Z_{PBL} \qquad (2)$$

where $z$ is the height, $L$ is the Monin-Obukhov length scale, $\phi_h(z) = \left(1 - 15 * \frac{z}{L}\right)^{-\frac{1}{2}}$, $\phi_m(z) = \left(1 - 15 * \frac{z}{L}\right)^{-\frac{1}{3}}$, $a$ = 0.85, and $k$ is the von Karman constant (= 0.4). Therefore, there is a discontinuity at the top of the surface layer ($z = 0.1 * Z_{PBL}$), where the Pr drops from a constant value $\mathrm{Pr}_{ML}$ in the mixed layer to $\mathrm{Pr}_{ML} - 0.34$ at the top of the surface layer and then increases towards the surface. At the same height, the discontinuity occurs in the $K_h$ profile, which is more obvious in the higher vertical resolution set (Figs. 7b and 8b). This induces the fake inversion in the potential temperature profile, which was mentioned in the previous paragraphs (Fig. 4c).

Figures 9a and 9b show TKE vertical profiles from the 18-level SCM simulations using UW01 and UW09 schemes and LES simulation. The LES model results show that the high TKE values appear in both the surface layer and mid-PBL, and the TKE value rapidly decreases due to the stable stratification near the PBL top. In the UW01 and UW09 schemes, the high TKE values in the suface layer cannot be captured. Above the surface layer, the TKE profiles in two schemes have a similar shape to that of the LES, however, the magnitudes are underestimated with biases of approximately 30%. In these two schemes, the minimum TKE above the PBL is zero, which is approximately 0.3 $m^2 s^{-2}$ in the LES. After raising the vertical resolution, the TKE bias in the UW09 scheme is greatly reduced, while that in the UW01 scheme shows little change (Fig. 9c). These features can also be clearly seen from the time evolution figure (Fig. 10). Similar to the $K_h$, as the PBL top raises, the maximum values of TKE generally increase, and the values greater than a certain small TKE (e.g., 0.5 $m^2 s^{-2}$) extend to higher levels. The TKE biases change little over time and are 1~1.5 $m^2 s^{-2}$ in the mixed layer.

**4. Summary**

An SCM based on parameterizations and the dynamic core inherited from the RegCM4 was successfully constructed. With the LES benchmark simulation results, the SCM model was tested in DCBL simulations. Despite the successful general DCBL simulations, discrepancies within individual SCM simulations do exist. Specifically, four PBL schemes (two of which were added into the SCM in this study) were further compared in terms of their performances for the PBL height, mixing strength, and vertical profiles of potential temperature, wind speed, sensible heat flux, eddy diffusivity, and TKE. The vertical resolution effect on the simulations was also discussed.

The diagnosed PBL height directly from the SCM is quite different among the four schemes due to the use of different calculation methods, which should not be used alone as an evaluation indicator. However, the PBL height can aid in indicating the time evolution of the entrainment zone height and vertical profiles of the heat flux, eddy diffusivity, and TKE. For the re-diagnosed PBL height using the same method, there is little difference among the schemes except the HB, and the biases relative to the LES result are small, indicating a successful general simulation with the DCBL.

In general, the YSU performs best in reproducing the LES results on nearly all variables evaluated, but the YSU still has considerable room for improvement. The major bias is that the wind speed simulated in the YSU is not as well mixed as that in the LES, which is also a common problem in all four schemes. The UW09 scheme has the second best performance. The wind speed simulated in the UW09 below the boundary layer top is slightly underestimated. The UW09 also underesimates the TKE, entrainment flux, and height of the entrainment zone. The UW01 ranks third, as the biases are similar to those of the UW09 but those of the UW01 are larger, and the simulated potential temperature is underestimated and not well mixed in the UW01.

The HB is the worst scheme. The major biases include the following: (1) The PBL height, entrainment flux, and height of the entrainment zone are overestimated. (2) The vertical gradients of the potential temperature and wind speed within the mixed layer are largely overestimated. (3) Due to a deficient paramerization of Pr, there is a fake inversion layer near the top of the surface layer.

Raising the vertical resolution in the lower atmosphere aids in simulation of the potential temperature and sensible heat flux profiles for all the schemes. However, considering the simulation of wind profiles, a higher vertical resolution is beneficial only for the YSU and UW09 schemes. In the 1.5-order closure schemes, UW01 and UW09, the TKE are calculated. In both two schemes, the TKE values within the PBL are underestimated in comparison with the LES model. Raising the vertical resolution helps to reduce the bias only in the UW09 scheme.

Notably, the assessments in this study are focused only on a single DCBL case. The comprehensive performance assessment on of a PBL scheme needs more cases, such as the cases of a stably stratified boundary layer, nocturnal stratocumulus-topped boundary layer, and real cases, which should be further studied in future works. Studies on the interaction between the PBL and moist convections, radiation processes, or surface processes are also desirable. Although further testing is needed, current simulations show encouraging results towards the use of SCM for the study of physical processes in RegCM4.

## Acknowledgements

This study was jointly supported by the National Key R&D Program of China (2018YFA0606301 and 2016YFC0402405) and the National Natural Science Foundation of China (41405101). Figures were prepared using the NCAR Command Language (NCL 2017). The SCM and LES results are available upon request from the corresponding author.

**Captions**

Table 1. Information regarding some RCMs and LAMs with SCM implementation.

Table 2. Comparisons of the thickness of well-mixed layer ($H_{ML}$) and mixing index (MI) caculated based on the hourly mean profiles. The values with the two smallest biases in a column are bolded, and the value with the smallest bias is also marked with an asterisk.

Fig. 1. Time series of simulated surface frictional velocity (m $s^{-1}$): (a) 18-level runs and (b) 41-level runs. The simulated results are from the LES (black), HB/HB_z41 (golden), UW01/UW01_z41 (red), YSU/YSU_z41 (blue), and UW09/UW09_z41 (green) experiments.

Fig. 2. Time series of simulated PBL height (m): (a, c) 18-level runs and (b, d) 41-level runs. The PBL heights are diagnosed using two methods: (a, b) output directly from respective schemes and LES and (c, d) re-diagnosed using the bulk Richardson number method.

Fig. 3. Simulated wind speed (units: m $s^{-1}$) profiles at (a) 5 h with the 18-level set, (b) 9 h with the 18-level set, and (c) 9 h with the 41-level set. In all panels, the horizontal lines denote the boundary layer top, which is output directly from the respective SCM schemes and LES. In both (a) and (b), the horizontal green, blue, and red lines are overlapped.

Fig. 4. Simulated potential temperature (units: K) profiles at (a) 5 h with the 18-level set, (b) 9 h with the 18-level set, and (c) 9 h with the 41-level set. In all panels, the horizontal lines denote the boundary layer top, which is output directly from the respective SCM schemes and LES. In both (a) and (b), the horizontal green, blue, and red lines are overlapped.

Fig. 5. Simulated vertical sensible heat flux (normalized by the surface flux) profiles at (a) 5 h with the 18-level set, (b) 9 h with the 18-level set, and (c) 9 h with the 41-level set. In all panels, the horizontal lines denote the boundary layer top, which is output directly from the respective SCM schemes and LES. In both (a) and (b), the horizontal green, blue, and red lines are overlapped.

Fig. 6. Time evolution of the vertical profiles of sensible heat fluxes (units: W $m^{-2}$) from (a) HB, (b) HB_z41, (c) UW01, (d) UW01_z41, (e) YSU, (f) YSU_z41, (g) UW09, (h) UW09_z41, and (i) LES. White lines denote the boundary layer top, which is output directly from the respective SCM schemes and LES run.

Fig. 7. Simulated (a, b) eddy diffusivity profiles for heat (units: $m^2$ $s^{-1}$) and (c, d) the Prandtl number at 5 h with (a, c) the 18-level set and (b, d) the 41-level set. In all panels, the horizontal lines denote the boundary layer top, which is output directly from the respective SCM schemes. In both (a) and (c), the horizontal green and red lines are overlapped.

Fig. 8. Time evolution of the vertical profiles of eddy diffusivity for heat (units: $m^2$ $s^{-1}$) from (a) HB, (b) HB_z41, (c) UW01, (d) UW01_z41, (e) YSU, (f) YSU_z41, (g) UW09, and (h) UW09_z41. White lines denote the boundary layer top, which is output directly from the respective SCM schemes.

Fig. 9. Simulated TKE (units: $m^2$ $s^{-2}$) at (a) 5 h with the 18-level set, (b) 9 h with the 18-level set, and (c) 9 h with the 41-level set. Horizontal lines denote the boundary layer top, which is output directly from respective SCM schemes and LES. In both (a) and (b), the horizontal green and red lines are overlapped.

Fig. 10. Time evolution of the vertical profiles of TKE (units: $m^2\ s^{-2}$) from (a) UW01, (b) UW01_z41, (c) UW09, (d) UW09_z41, and (e) LES. White lines denote the boundary layer top, which is output directly from the respective SCM schemes.

Table 1. Information regarding some RCMs and LAMs with SCM implementation.

| RCM/ LAM | ARCSyM | GRAPES_Meso | GRIMs | HIRLAM/HARMONIE | MM5 | WRF |
|---|---|---|---|---|---|---|
| Institution | University of Colorado, US | China Meteorological Administration, China | Yonsei University, South Korea | Several National Meteorological Services in Europe | Pennsylvania State University and NCAR, US | NCAR, US |
| Reference | Morrison et al. 2003 | Yang and Shen 2011 | Hong et al. 2013 | Neggers et al. 2012 | Deng et al. 2003 | Hacker and Angevine 2013 |

Table 2. Comparisons of the thickness of well-mixed layer ($H_{ML}$) and mixing index (MI) caculated based on the hourly mean profiles. The values with the two smallest biases in a column are bolded, and the value with the smallest bias is also marked with an asterisk.

| Experiment | $H_{ML}$ (m) | | MI (0.1 K $km^{-1}$) | |
|---|---|---|---|---|
| | 5h | 9h | 5h | 9h |
| LES | 1218.9 | 1970.2 | 0.11 | 0.08 |
| HB/HB_z41 | 793.5/714.0 | **1586.3***/1752.1 | 1.27/0.86 | **0.27**/0.34 |
| YSU/YSU_z41 | **965.8*/1422.8** | **1473.2/2135.6** | **0.58*/0.18*** | **0.14*/0.12*** |
| UW01/UW01_z41 | 696.3/794.6 | 1215.7/1629.3 | 0.82/0.45 | 0.52/0.36 |
| UW09/UW09_z41 | **800.8/1281.1*** | 1343.3/**2017.5*** | **0.79/0.28** | 0.52/**0.19** |

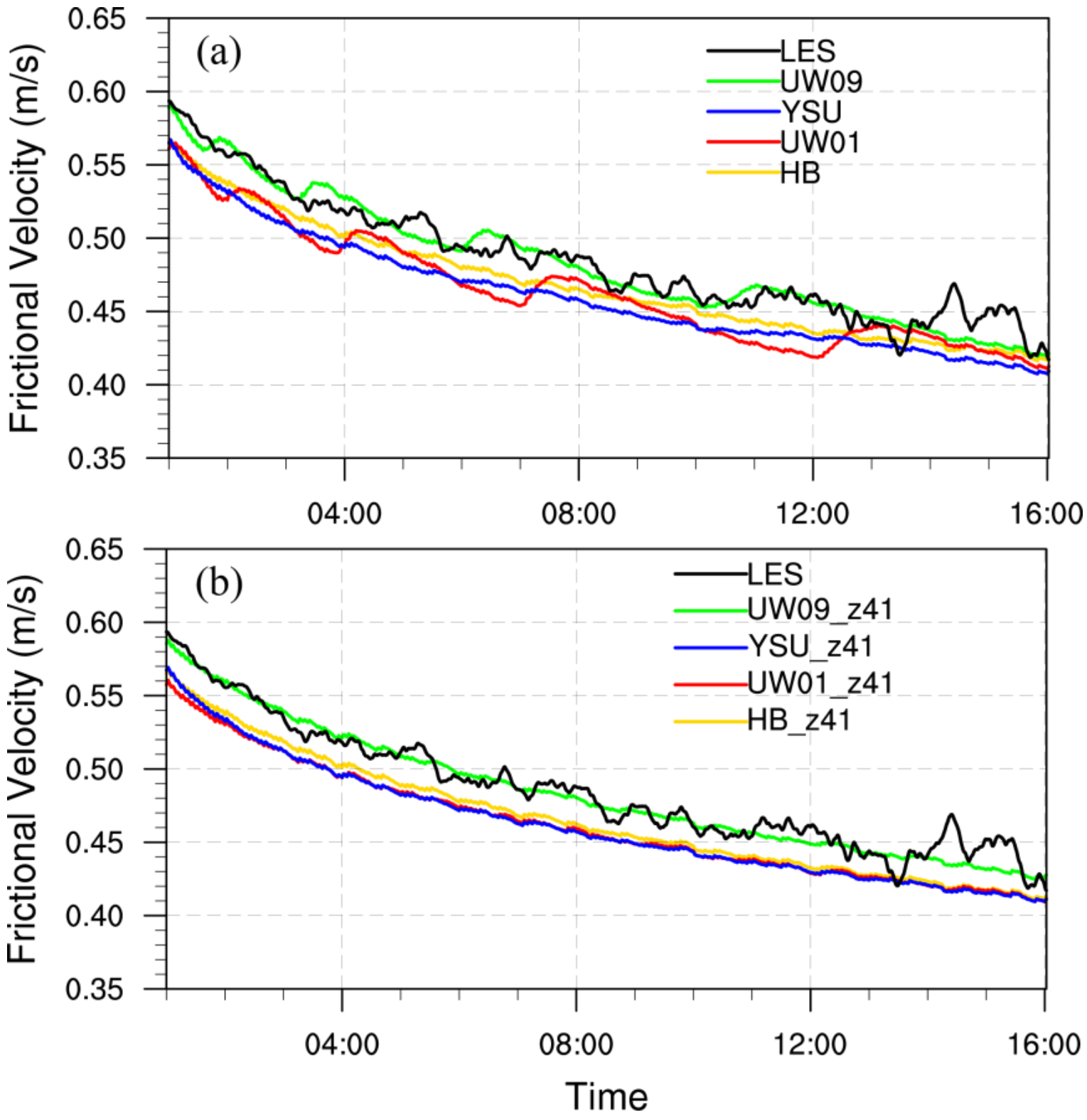

Fig. 1. Time series of simulated surface frictional velocity (m $s^{-1}$): (a) 18-level runs and (b) 41-level runs. The simulated results are from the LES (black), HB/HB_z41 (golden), UW01/UW01_z41 (red), YSU/YSU_z41 (blue), and UW09/UW09_z41 (green) experiments.

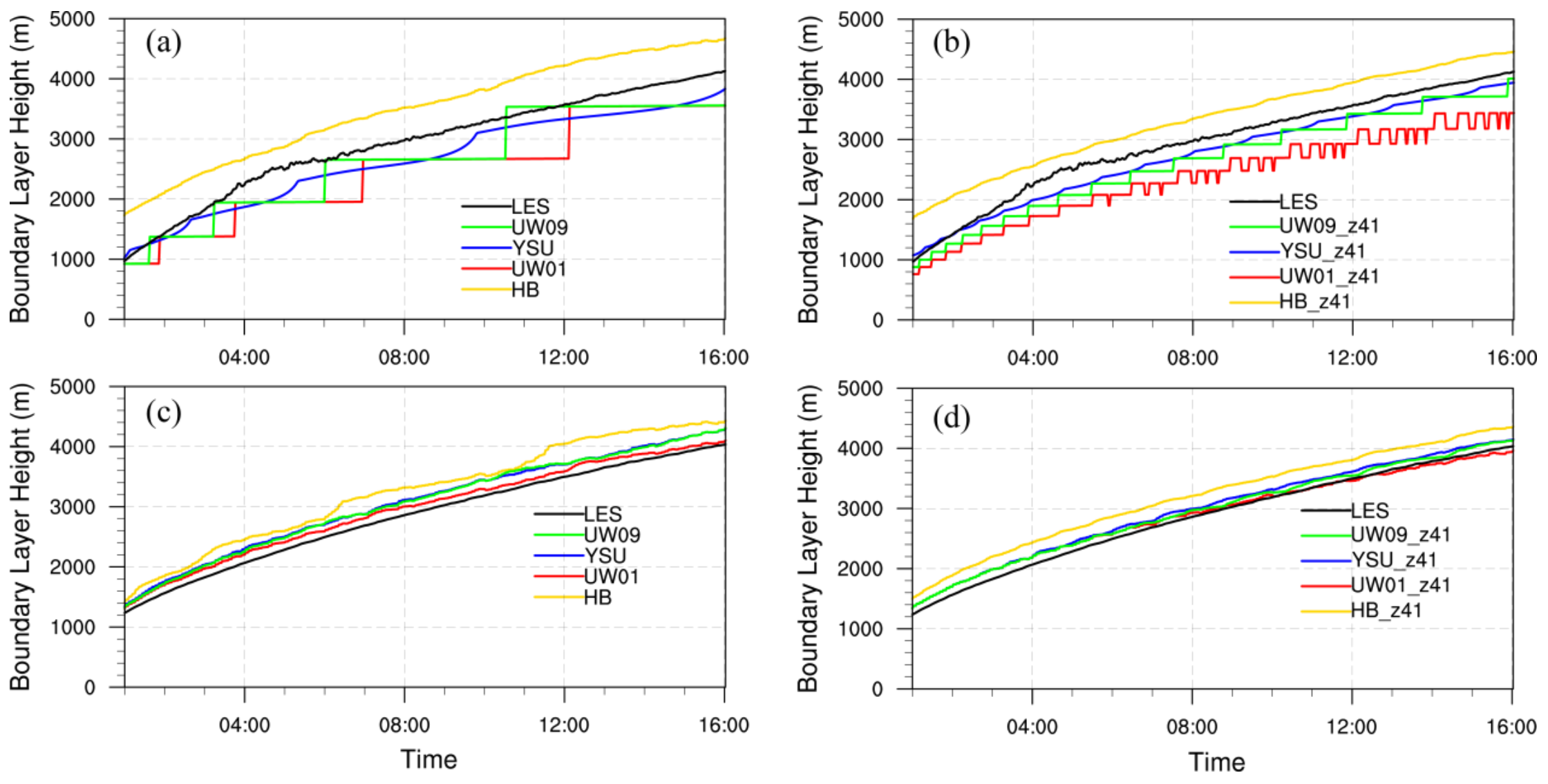

Fig. 2. Time series of simulated PBL height (m): (a, c) 18-level runs and (b, d) 41-level runs. The PBL heights are diagnosed using two methods: (a, b) output directly from respective schemes and LES and (c, d) re-diagnosed using the bulk Richardson number method.

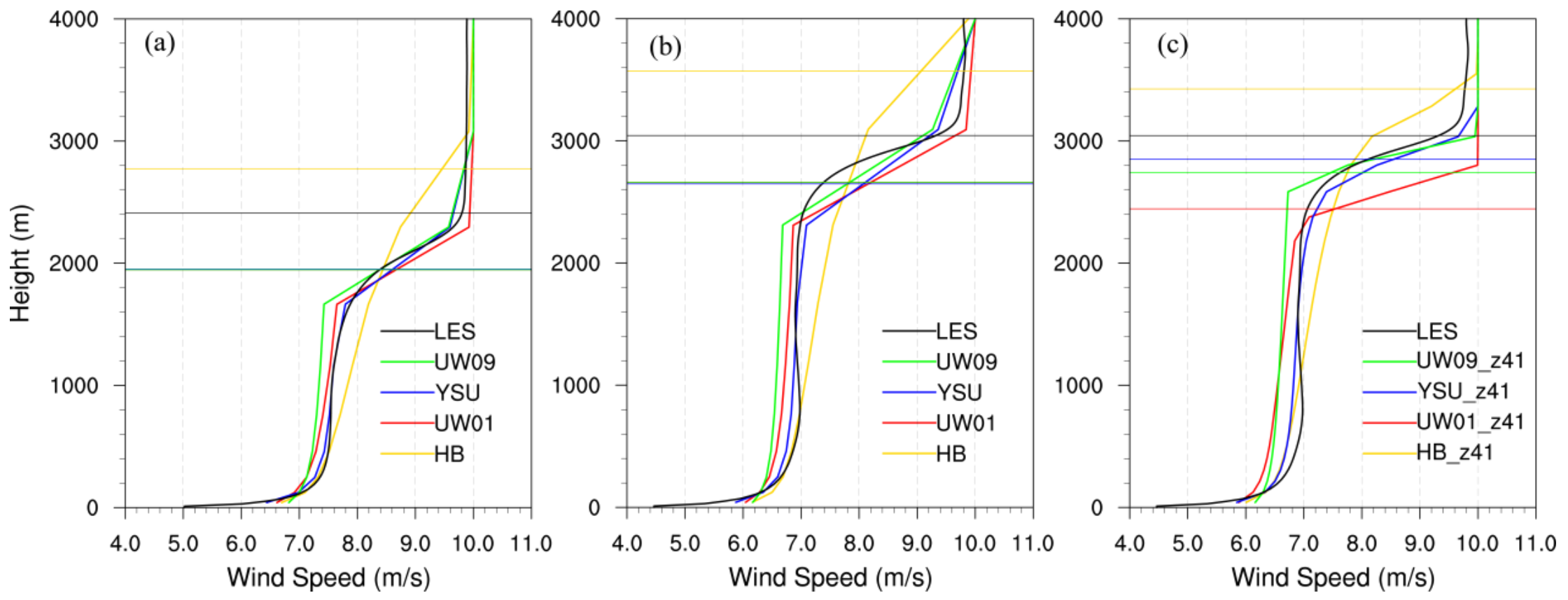

Fig. 3. Simulated wind speed (units: m $s^{-1}$) profiles at (a) 5 h with the 18-level set, (b) 9 h with the 18-level set, and (c) 9 h with the 41-level set. In all panels, the horizontal lines denote the boundary layer top, which is output directly from the respective SCM schemes and LES. In both (a) and (b), the horizontal green, blue, and red lines are overlapped.

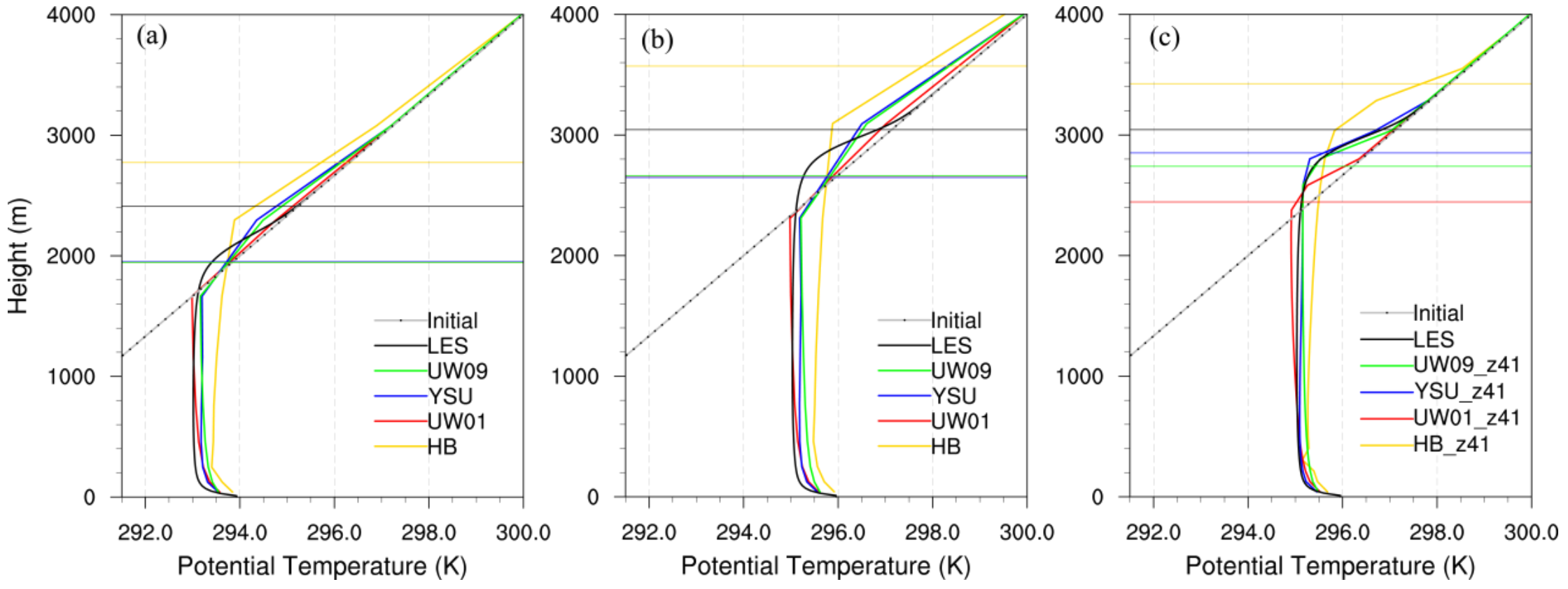

Fig. 4. Simulated potential temperature (units: K) profiles at (a) 5 h with the 18-level set, (b) 9 h with the 18-level set, and (c) 9 h with the 41-level set. In all panels, the horizontal lines denote the boundary layer top, which is output directly from the respective SCM schemes and LES. In both (a) and (b), the horizontal green, blue, and red lines are overlapped.

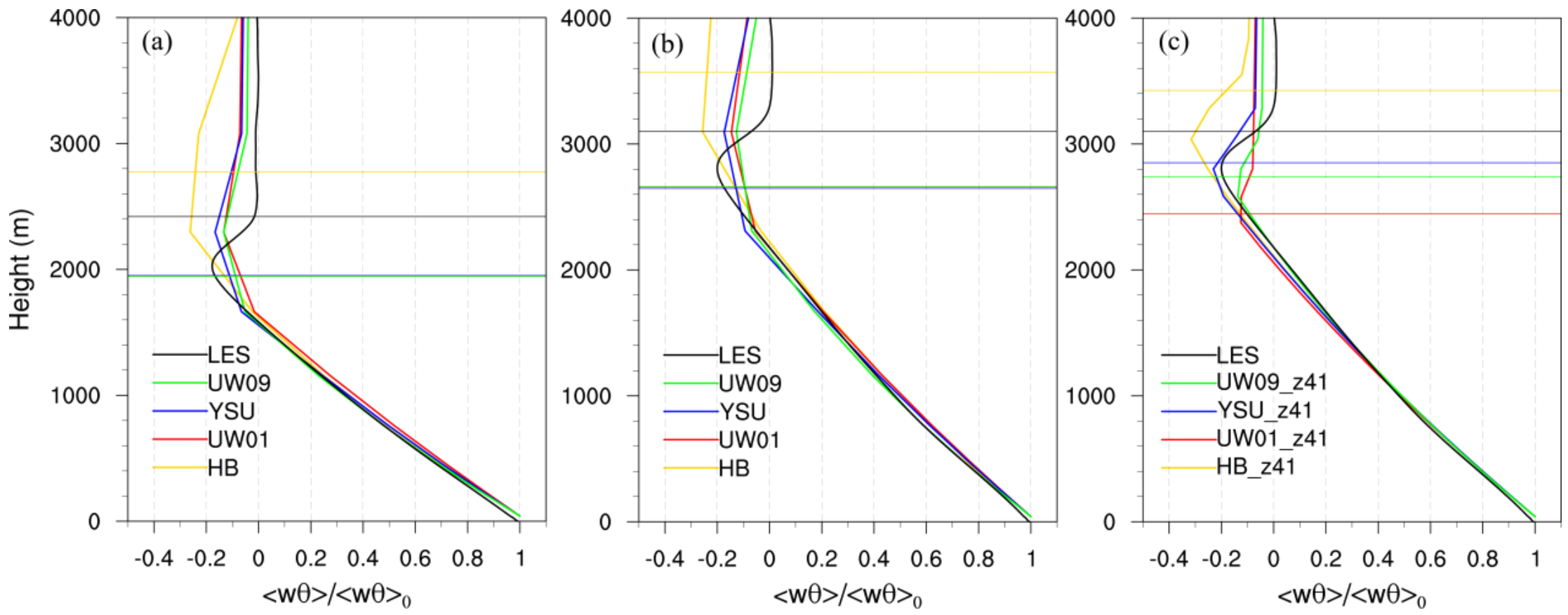

Fig. 5. Simulated vertical sensible heat flux (normalized by the surface flux) profiles at (a) 5 h with the 18-level set, (b) 9 h with the 18-level set, and (c) 9 h with the 41-level set. In all panels, the horizontal lines denote the boundary layer top, which is output directly from the respective SCM schemes and LES. In both (a) and (b), the horizontal green, blue, and red lines are overlapped.

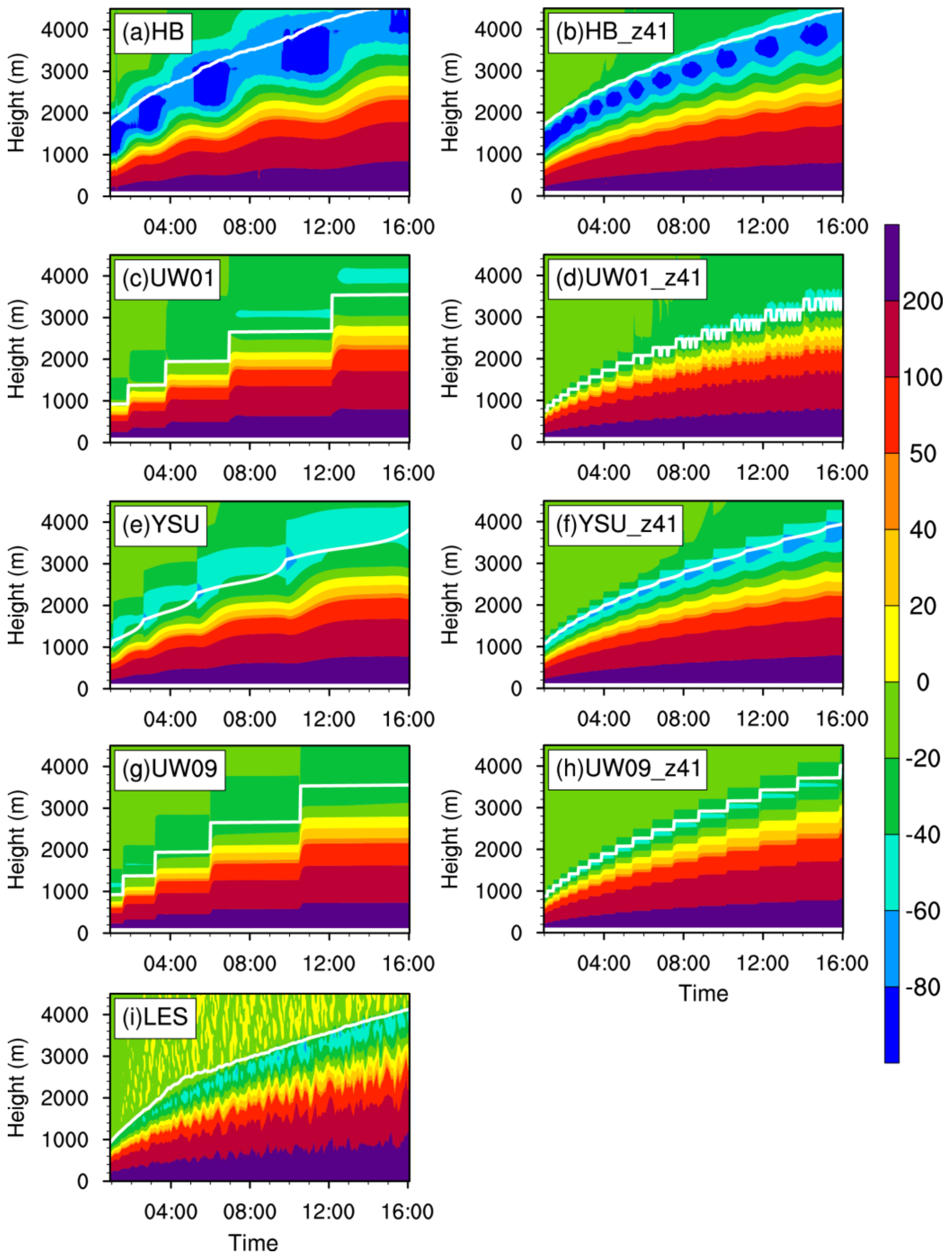

Fig. 6. Time evolution of the vertical profiles of sensible heat fluxes (units: W m$^{-2}$) from (a) HB, (b) HB_z41, (c) UW01, (d) UW01_z41, (e) YSU, (f) YSU_z41, (g) UW09, (h) UW09_z41, and (i) LES. White lines denote the boundary layer top, which is output directly from the respective SCM schemes and LES run.

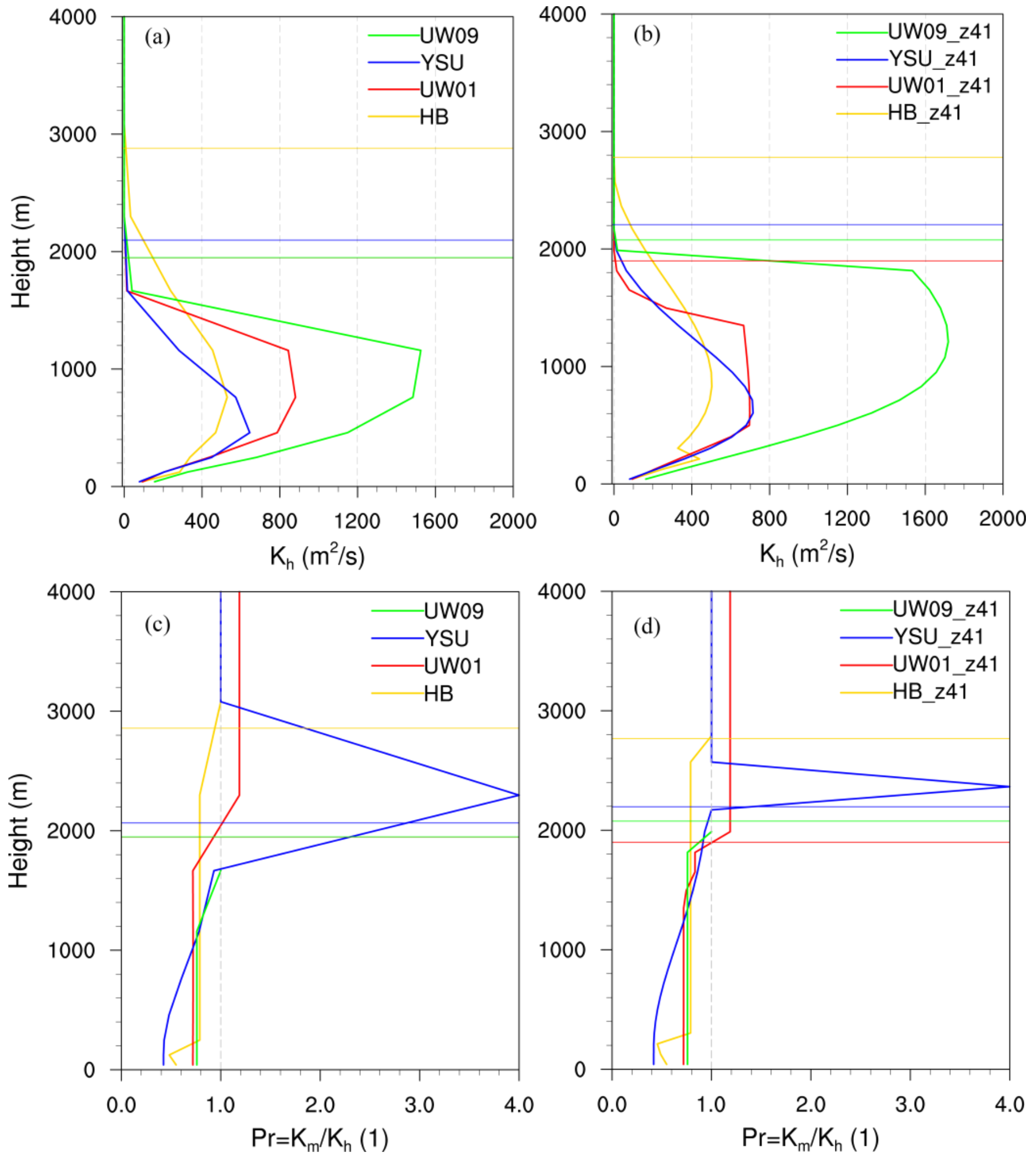

Fig. 7. Simulated (a, b) eddy diffusivity profiles for heat (units: $m^2 s^{-1}$) and (c, d) the Prandtl number at 5 h with (a, c) the 18-level set and (b, d) the 41-level set. In all panels, the horizontal lines denote the boundary layer top, which is output directly from the respective SCM schemes. In both (a) and (c), the horizontal green and red lines are overlapped.

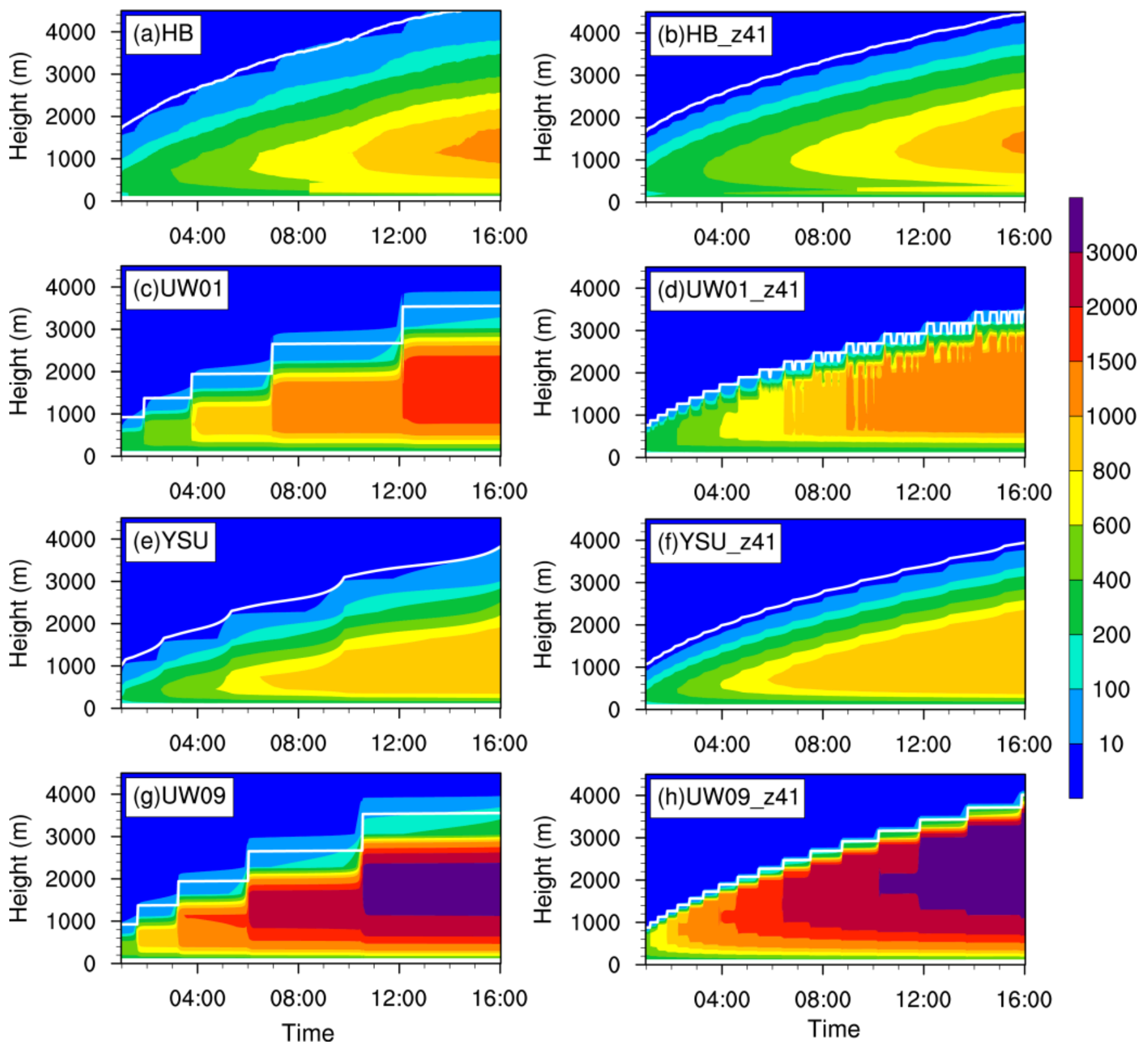

Fig. 8. Time evolution of the vertical profiles of eddy diffusivity for heat (units: $m^2\ s^{-1}$) from (a) HB, (b) HB_z41, (c) UW01, (d) UW01_z41, (e) YSU, (f) YSU_z41, (g) UW09, and (h) UW09_z41. White lines denote the boundary layer top, which is output directly from the respective SCM schemes.

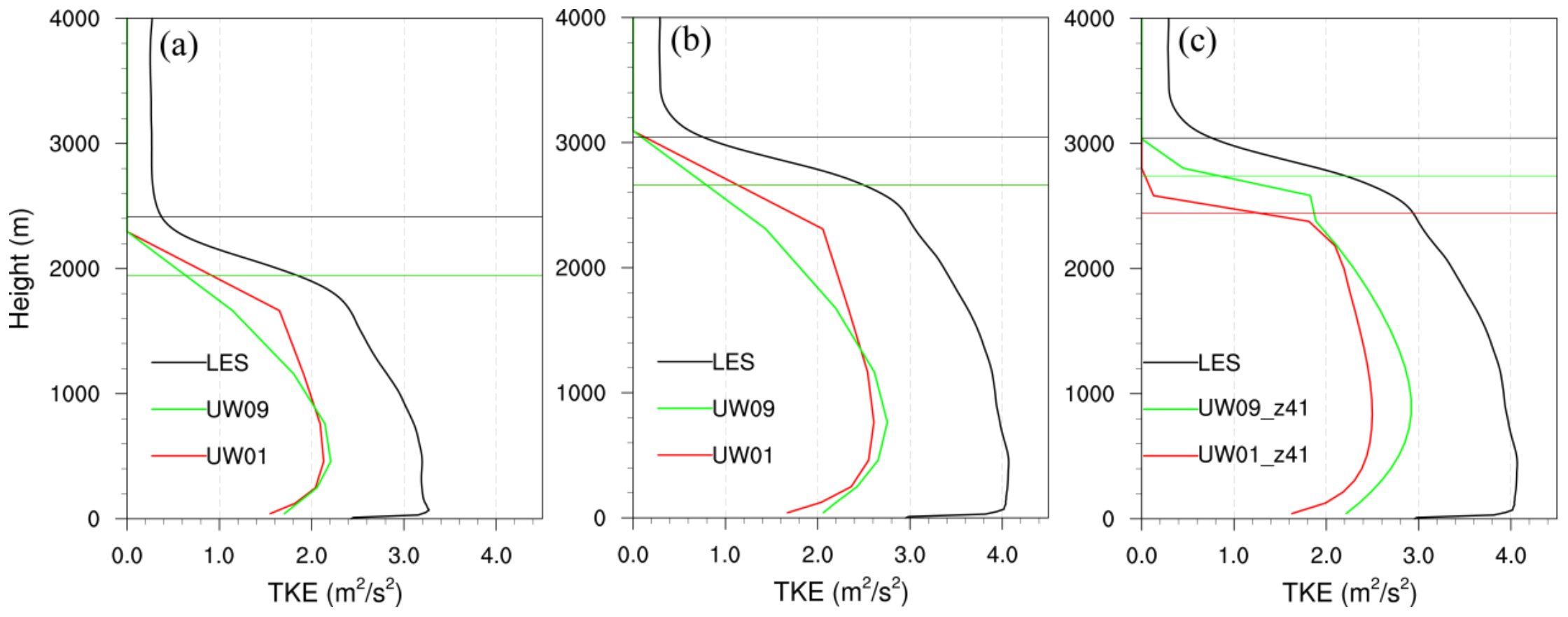

Fig. 9. Simulated TKE (units: $m^2$ $s^{-2}$) at (a) 5 h with the 18-level set, (b) 9 h with the 18-level set, and (c) 9 h with the 41-level set. Horizontal lines denote the boundary layer top, which is output directly from respective SCM schemes and LES. In both (a) and (b), the horizontal green and red lines are overlapped.

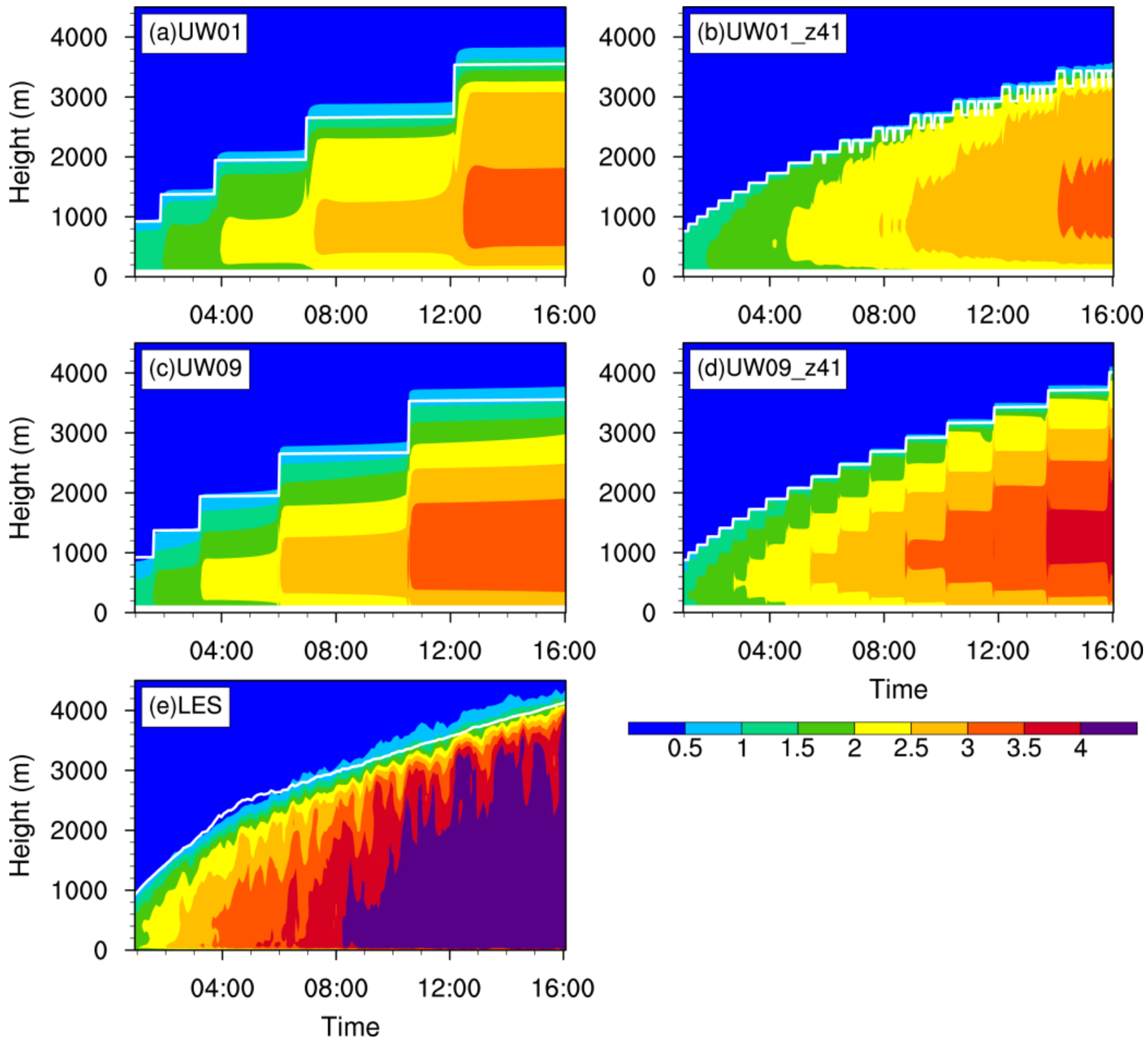

Fig. 10. Time evolution of the vertical profiles of TKE (units: $m^2 s^{-2}$) from (a) UW01, (b) UW01_z41, (c) UW09, (d) UW09_z41, and (e) LES. White lines denote the boundary layer top, which is output directly from the respective SCM schemes.